\documentclass[prb,twocolumn,showpacs,amsmath,amssymb,superscriptaddress]{revtex4}

\usepackage{graphicx}
\usepackage{amssymb,amsmath}
\usepackage{dcolumn}
\usepackage{bm}
\usepackage{color}

\begin{document}

\title{Multistability of cavity exciton-polaritons affected by the thermally generated exciton reservoir}

\author{D.V. Vishnevsky}
\affiliation{LASMEA, Nanostructure and Nanophotonics group, Clermont Universit\'{e}, Universit\'{e} Blaise Pascal, CNRS, 63177 Aubi\`{e}re Cedex France}

\author{D.D. Solnyshkov}
\affiliation{LASMEA, Nanostructure and Nanophotonics group, Clermont Universit\'{e}, Universit\'{e} Blaise Pascal, CNRS, 63177 Aubi\`{e}re Cedex France}

\author{N.A. Gippius}
\affiliation{LASMEA, Nanostructure and Nanophotonics group,
Clermont Universit\'{e}, Universit\'{e} Blaise Pascal, CNRS, 63177 Aubi\`{e}re Cedex France}
\affiliation{A. M. Prokhorov General Physics Institute, RAS, Vavilova Street 38, Moscow 119991, Russia}

\author{G. Malpuech}
\affiliation{LASMEA, Nanostructure and Nanophotonics group, Clermont Universit\'{e}, Universit\'{e} Blaise Pascal, CNRS, 63177 Aubi\`{e}re Cedex France}

\begin{abstract}
Until now, the generation of an excitonic reservoir in a cavity polariton
system under quasi-resonant pumping has always been neglected. We show that in microcavities
having a small Rabi splitting (typically GaAs cavities with a single quantum well),
this reservoir can be efficiently populated by polariton-phonon scattering.
We consider the influence of the exciton reservoir on the energy shifts of the resonantly pumped  polariton modes.
We show that the presence of this reservoir effectively reduces the spin anisotropy of the polariton-polariton interaction, in agreement with
recent experimental measurements, where the multistability of cavity polaritons has been analyzed [\emph{Nature Materials} \textbf{9}, 655-660 (2010)].
\end{abstract}

\maketitle

\section{Introduction}

Exciton-polaritons (polaritons) are the quasiparticles formed of photons and excitons strongly coupled in microcavities\cite{Microcavities}. Their popularity comes from their unique properties, mixing those of light (small effective mass) and excitations of matter (self-interactions and thermalization with phonons). Polaritons are photonic particles interacting with each other via a well-defined microscopic mechanism: exciton-exciton interaction. As a result, strongly coupled microcavities show a very strong non-linear optical response. Parametric oscillations and amplification \cite{Savvidis, Stevenson, Ciuti} have been demonstrated a decade ago, showing a record efficiency, whereas low-threshold bistability has been demonstrated in 2004 \cite{Baas}. Polaritons also possess two spin projections which can lead to non-linear spin dependent effects, allowing to use microcavities as polarization-selective optoelectronic devices, or "spin-optronic" devices \cite{ReviewSpin}. The first method of controlling the polariton spin degree of freedom is the splitting existing between the TE and TM polarized polariton modes. The effect of this splitting can be mathematically described as an effective magnetic field acting on the polariton pseudospin. It is at the origin of time-dependent spin oscillations  \cite{KKavokin} and of the optical spin Hall effect \cite{AKavokin, Leyder}. The spin degree of freedom can be used to generate half-integer topological defects such as half-solitons \cite{FlayacHalfSoliton}.

Another key ingredient of the polariton spin dynamics is the spin anisotropy of the polariton-polariton interaction which was noticed for the first time in \cite{Shelykh2004}. Exciton-exciton interaction constant in the triplet configuration $\alpha_{1}$ is dominated by Coulomb exchange interaction. It can be calculated in the Born approximation similarly to the spin-less case \cite{Ciuti98}. On the other hand, an exchange process between excitons having opposite spins $\pm 1$ (singlet configuration) leads to the formation of dark excitons having a spin $\pm 2$. The polariton-polariton interaction in the singlet configuration is therefore a second-order process which invokes intermediate states (typically located half a Rabi splitting above the polariton state). The resulting interaction  is weakly attractive (constant $\alpha_{2}$), much weaker than the interaction in the triplet configuration \cite{Renucci2005, Glazov}. This spin anisotropy led to the prediction and observation of a wide variety of remarkable phenomena, such as the inversion of the linear polarization during polariton-polariton scattering processes \cite{Renucci2005}, the spin-Meissner effect \cite{Meissner}, the stability of half-topological defects \cite{Solnyshkov}, and the polariton multistability \cite{Gippius, Gavrilov},
which has been recently evidenced experimentally \cite{Gavrilov_res,Gavrilov_PRB,Sarkar, Paraiso}.
The surprising conclusion of the latter experiment\cite{Paraiso} was that the theoretical description based on the driven-dissipative Gross-Pitaevskii equation required to consider a repulsive polariton-polariton interaction in the singlet configuration in order to fit the experiments. This conclusion was also based on recent observations suggesting that the bi-exciton state could play the role of the intermediate state for the second order polariton-polariton interaction \cite{Masha}. As a result, $\alpha_{2}$ could become comparable with $\alpha_{1}$ and change sign when polariton energy becomes larger than the bi-exciton energy.

It is worth to mention here that a description of the resonantly pumped polariton system by the Gross-Pitaevskii equation, widely used nowadays, is based on the assumption that the majority of the polaritons occupies the coherent states. However, the systematic time-resolved transmission experiments in $ns$ time domain \cite{Gavrilov_res,Gavrilov_PRB} have clearly shown the important role of long-living states ($\sim 500$~ps lifetime) populated under resonant excitation. The experimentally observed blue shift of the transmitted light was shown to be different from the temporal profile of coherent polariton density. The blue shift remained of the order of meV well after the coherent polaritons have decayed. This type of effect cannot be described within the Gross-Pitaevskii approach, where the only source of energy shifts are the coherent polaritons. In order to explain these experimental findings, the authors have considered phenomenological coupling of the resonantly pumped polariton mode with a long-lived "reservoir" states. Although the exact nature of both "reservoir" and coupling  was not specified, the success in the description of the wide experimental data set (also for the case of $cw$  resonant pumping \cite{Sarkar}) has proven the important role of the "reservoir" in these experiments. This concept also allowed to resolve several contradictions in the interpretation of data obtained in $cw$ and ps-range pulsed excitation experiments (see detailed discussion in Ref.\cite{Gavrilov_PRB}). A coupling with the reservoir has also had to be introduced to describe OPO polarization dynamics in earlier experiments \cite{Solnyshkov2007}

\begin{figure}[h]
  \includegraphics[width=0.4\textwidth,clip]{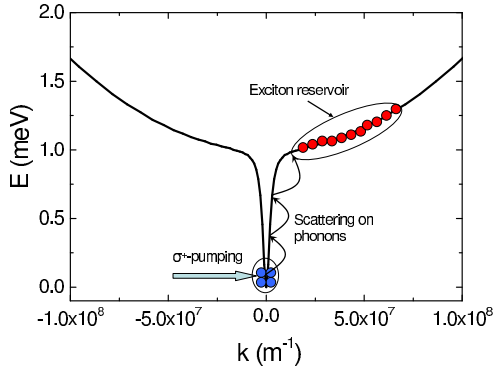}\\
  \caption{(Color online) The scheme of the process we considered. The circularly polarized pumping is in the ground state. The interaction with phonons provides an exchange of polaritons between the condensate and the reservoir.}
  \label{Fig1}
\end{figure}

In the present work, we propose to describe theoretically the thermal generation of an excitonic reservoir from a resonantly pumped low energy polariton state. To our knowledge, this effect has never been studied, being always considered as negligible. As we are going to show below, this is essentially incorrect in structures showing a small Rabi splitting at zero or positive detunings, namely in the widely-studied GaAs-based cavities containing a single QW. This is the type of structure which has been used so far in multistability experiments \cite{Sarkar,Paraiso}. In the first part of the paper we solve the semi-classical Boltzmann equations with the ground state being continuously pumped by a resonant, circularly polarized ($\sigma^{+}$) laser as sketched on the figure 1. This way of using Boltzmann equations is extremely unusual for the field of exciton-polaritons, since such equations are typically used in order to describe the non-resonant pumping case and the \emph{relaxation} of reservoir excitons toward the lower energy states \cite{Microcavities}. We show that in the above-mentioned type of structures, the resonant pumping results in the formation of a thermal exciton reservoir of moderate density which we assume to be unpolarized whatever the polarization of the pumping laser. This assumption might not be fully correct, since the interactions with the strongly polarized pumped should split the circularly polarized reservoir state, which might slow down the standard exciton spin relaxation mechanisms \cite{Maialle}. In the absence of any data allowing to verify the presence of such effect, we are going to assume throughout this paper that the exciton spin relaxation is much faster than the reservoir exciton lifetime, leading to an unpolarised reservoir. 
This reservoir can provoke a blue shift of both polariton spin components 
by a fraction of meV, comparable with the energy shifts induced by the resonantly pumped state occupation. This can result in an apparent positive $\alpha_{2}$, induced by the reservoir. In the second part of the paper we solve self-consistently the coupled spinor Gross-Pitaevskii--Boltzmann equations, using the proper attractive $\alpha_{2}$ constant for polariton-polariton interaction. We show that the correct accounting for the reservoir generation by acoustic phonons is sufficient to describe the most important features of the experiment of Ref.\cite{Paraiso}.

\section{Analytical approximation}

We begin with a simple analytical model allowing to estimate the importance of the effect we would like to discuss, that is, the population of an exciton reservoir by scattering with phonons up from the coherently pumped condensate. The spatially homogeneous condensate and the reservoir are described by the occupation numbers. In what follows we will use lowercase letters to describe the surface densities (e.g. the condensate density $n_0=N_0/S$) and uppercase ones for the occupancy of the states.

In the most general case the semi-classical Boltzmann equation for the occupation number of the state with wave-vector $\vec k$ reads \cite{CavityPolaritons}:
\begin{widetext}
\begin{equation}
\frac{{dN_{\vec k} }}{{dt}} = P_{\vec k}  - \Gamma _{\vec k} N_{\vec k}  - N_{\vec k} \sum\limits_{\vec k'} {W_{\vec k \to \vec k'} } (1 + N_{\vec k'} ) + (1 + N_{\vec k} )\sum\limits_{\vec k'} {W_{\vec k' \to \vec k} } N_{\vec k'}
\label{BoltzmannEq}
\end{equation}
\end{widetext}

Here the terms with $W$ are the scattering rates between different states, $P$ is the pumping and $\Gamma$ are the decay rates of the corresponding states. The terms $(1 + N_{\vec k})$ account for the bosonic nature of the particles involved, allowing to describe bosonic stimulation towards strongly populated states.

This system of equations will be solved as a whole below, but first we will discuss an approximated solution allowing analytical treatment. Indeed, if one assumes the conditions when single-phonon assisted scattering is stronger than multi-scattering channels, the reservoir states become only coupled with the ground state, and not between each other. In QW-based structures, the exciton-phonon coupling arises between 2D excitons and 3D phonons. The wave vector in the plane is therefore conserved, which is not the case for the wave vector of phonons in the $z$ direction ($q_z$). This specificity allows to organize energy-conserving scattering processes even in the very sharp polariton dispersion. This process is however limited by a wave vector cutoff provided by $q_{z,max} \sim  1/L$, $L$ being the well width. In practice this cutoff limits the energy exchanged during a polariton-acoustic phonon scattering event to 2-3 meV.
So qualitatively, one expects single-scattering processes to be dominant if the energy difference between polariton ground state and the excitonic reservoir is smaller than 3 meV, and multi-scattering processes if this energy difference is larger than 3 meV.

In the framework of the above-mentioned assumptions, the equation of motion for the reservoir state $\vec k$ writes:
\begin{widetext}
\begin{equation}
\frac{{dN_{\vec k} }}{{dt}} =  - \Gamma _{\vec k} N_{\vec k}  - N_{\vec k} W_{\vec k \to 0} (1 + N_0 ) + (1 + N_{\vec k} )W_{0 \to \vec k} N_0
\label{KineticsK}
\end{equation}
\end{widetext}

Using the Fermi golden rule and assuming Lorentzian broadening of polariton lines, one can calculate the scattering rate for polariton-acoustic phonon interaction:
\begin{widetext}
\begin{equation}
W_{\vec k \to \vec k'}^{phon}  = \frac{{2\pi }}{\hbar }\sum\limits_{{\vec q}_z} {|M(\vec q)|} ^2 (0,1 + N^{\vec q = \vec k - \vec k' + \vec q_z }_{ph} ) \times \frac{{\hbar \gamma _{\vec k'} /\pi }}{{(E(k') - E(k) \pm \hbar \omega _{\vec q} )^2  + (\hbar \gamma _{k'} )^2 }}
\label{ScatRate}
\end{equation}
\end{widetext}

Here $\gamma_k$ is the broadening of the polariton line corresponding to the decay rate $\Gamma_{\vec k}$ in the Boltzmann equations, $\omega_q$ - the phonon frequency, $N^{\vec q = \vec k - \vec k' + \vec q_z }_{ph}$ - number of phonons, $M(q)$ - the matrix element of polariton-phonon interaction, taking into account the excitonic fraction of each of the states involved.
In order to further simplify this equation, we assume that the only dependence of $W_{\vec k \to \vec k'}^{phon}$ between the ground state and the reservoir state is
via the number of phonons $N_{ph}^{\vec k}$: $W_{0 \to \vec k} = W N_{ph}^{\vec k}$ and $W_{\vec k \to 0} = W (N_{ph}^{\vec k}+1)$ where $N_{ph}^{\vec k}=1/(\exp ( - \Delta E_{k}/k_{B}T) - 1)$. We take the value of $W$ as a constant, which will be shown in section IV to possess a value of order $10^{7}s^{-1}$ using an exciton decay rate $\Gamma _k  \approx \hbar/400$~ps. Thus, the above-mentioned wavevector cutoff is neglected together with the wavevector dependence of the interaction with acoustic phonons, and the only dependence that is kept is the exponential decrease of phonon occupation numbers for large energies exchanged.

For this system of equations, one can analytically obtain the stationary values of the occupation numbers:

\begin{equation}
N_{\vec k}  = \frac{{N_{ph}^{\vec k} }}{{N_0  + N_{0}^{c}}}N_0
\label{NKStat}
\end{equation}

from which the total reservoir density can be straightforwardly obtained.
Here $N_{0}^{c}=N_{ph}^{\vec k}  + 1 + {\raise0.7ex\hbox{${\Gamma _{\vec k} }$} \!\mathord{\left/
 {\vphantom {{\Gamma _{\vec k} } W}}\right.\kern-\nulldelimiterspace}
\!\lower0.7ex\hbox{$W$}}$ - critical value for $N_0$, discussed below.
 
From the previous equation, one can see the existence of two distinct regimes. If $N_0$ is smaller than $N_{0}^{c}$, the reservoir density increases linearly with $N_0$. It then saturates when $N_{0} \gg N_{0}^{c}$.
In the latter regime, the reservoir density does not depend on pumping anymore and is proportional to the phonon distribution function. In this regime, the reservoir distribution function is therefore exactly by a Bose distribution function with chemical potential equal $0$, if the energy is counted from the resonantly pumped polariton mode.

The results of the calculations are presented on the figure 2. The dashed lines show the reservoir density $n_{r}$ calculated with the procedure described above in the saturated regime versus the exciton-photon detuning for 3 different temperatures. The Rabi splitting of the structure is 4 meV, which corresponds to the case of a GaAs cavity containing a single QW \cite{Sarkar, Paraiso}. Going towards positive detuning reduces the energy splitting between the ground state and the reservoir, which increases the number of phonons. One can also increase the number of phonons directly, by simply increasing the temperature. In both cases, the reservoir density increases.

Next, in order to check the validity range of our analytical approximation, we have solved numerically the whole system of semi-classical Boltzmann equations where all possible scattering paths are taken into account Eq.\ref{BoltzmannEq}. The pumping term is different from zero only for ground state $P_{0}  \ne 0$. The simulation is carried out until the system reaches a steady-state distribution $N_{\vec k}$. 

The results of the numerical simulations are presented on the figure 2 with solid curves. As expected, the analytical approach shows a good accuracy in the positive detuning range when the energy difference between the polariton ground state and the excitonic reservoir is small. In that range, and at 20 K, the densities found are likely to destroy the strong coupling regime, and anyway to provoke strong blue shift due to exciton-exciton interactions. By going to negative detuning, the generated reservoir density  drops very rapidly, becoming completely negligible. This drop is explained by the fast decay of the direct scattering processes and the inefficiency of the multiple scattering channels. One should also note that the accounting for polariton-polariton scattering mechanisms brings no change to the present picture as soon as only the polariton ground state is pumped.

 \begin{figure}[h]
  \includegraphics[width=0.5\textwidth,clip]{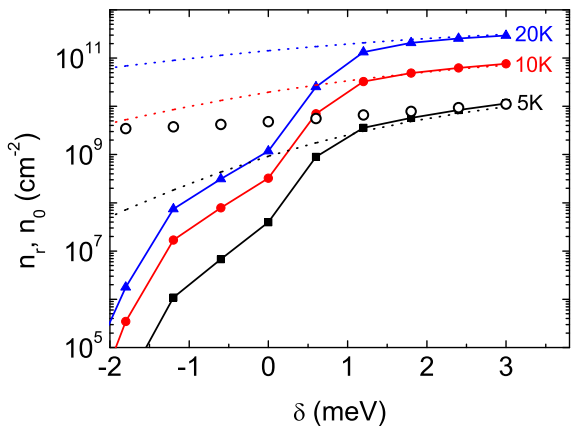}\\
  \caption{(Color online) Densities of excitons versus detuning for different values of temperature, solid lines - numerically calculated, dashed lines - analytical approximation. Open circles - density of condensate $n_{0}$. Rabi splitting 4 meV. }
  \label{Fig2}
\end{figure}

\begin{figure}[h]
  \includegraphics[width=0.5\textwidth,clip]{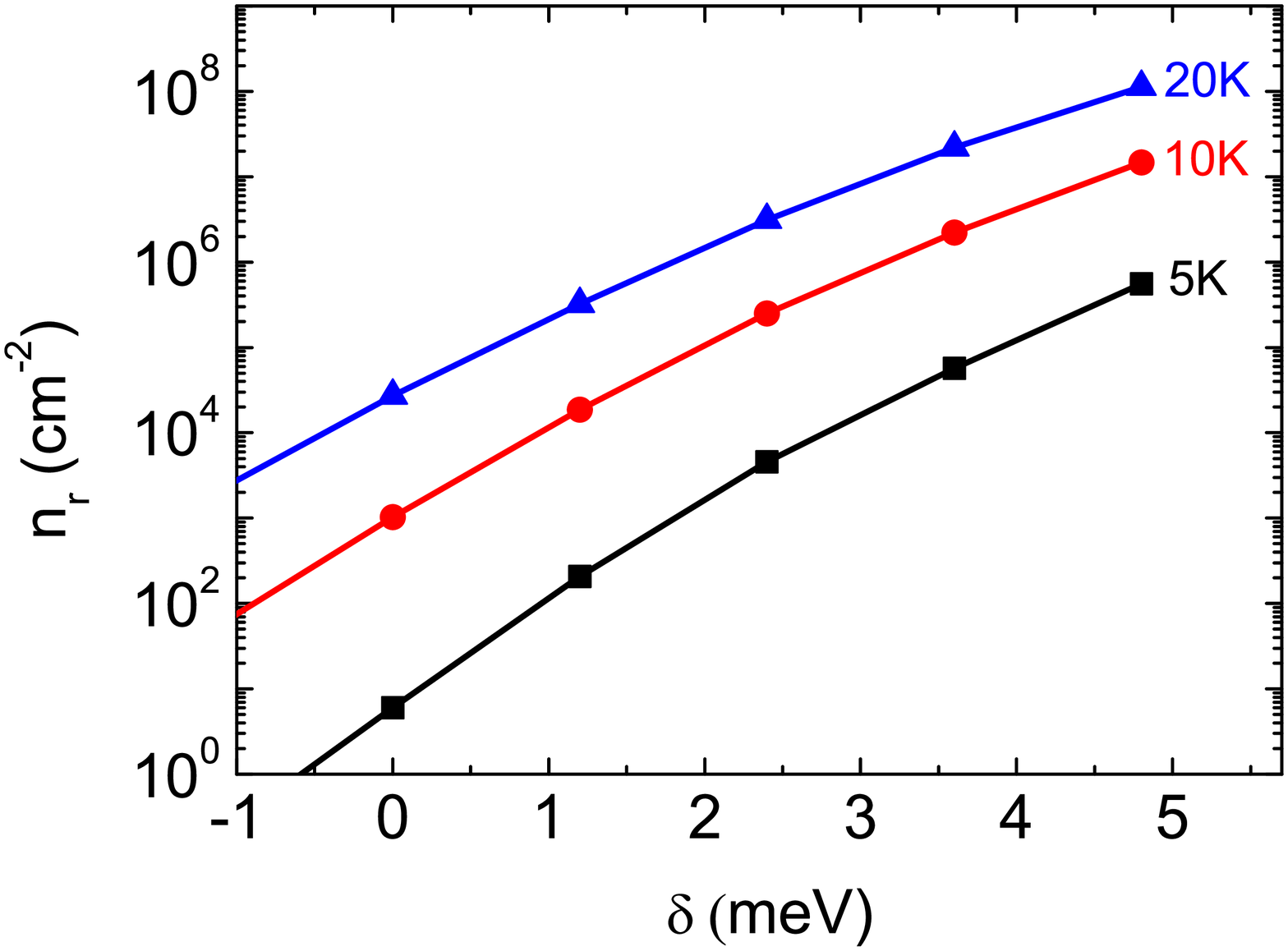}\\
  \caption{(Color online) Densities of excitons versus detuning for different values of temperature. Rabi splitting 12 meV }
  \label{Fig3}
\end{figure}

The simulations using the complete system of Boltzmann equations have also been performed for a GaAs microcavity showing a significantly larger Rabi splitting of 12 meV. The results are presented on the  Fig. \ref{Fig3}. In that case, the reservoir density generated by the resonant pumping is completely negligible, keeping below to $10^8$ cm$^{-2}$ for the values of detuning considered, and the analytical approximation of single-phonon scattering is not applicable.

\section{Influence of the reservoir on energy shifting}

Our next step is now to consider the impact of the excitonic reservoir, generated under quasi-resonant pumping, on the energy of the two spin components of the condensate. The interaction between excitons has been previously studied theoretically and experimentally, both in the scalar\cite{Ciuti98} and spinor cases\cite{Renucci2005,Masha}, taking the account the fact that the excitons are composite bosons \cite{Glazov}. We assume that this interaction is the same, whether the excitons concerned are both in the condensate, or one of them is in the reservoir.

In the triplet configuration the 2D interaction constant comes mostly from the exchange Coulomb interaction. It can be calculated in the Born approximation yielding the approximate formula $\alpha_{1}=3 E_{b}a_{B}^{2}$. In the singlet configuration, the exciton-exciton interaction is a second-order process passing either through the dark exciton states possessing spin $\pm 2$ or trough bi-exciton states. These latter states are usually several meV above the polariton states, which has two consequences: the matrix element of interaction becomes reduced and obtains a negative sign, as any second-order correction for the ground state. We should note, however, that it is possible to have the biexciton state close to the polariton ground state, if the Rabi splitting is small and the detuning is strongly positive\cite{Masha}. In what follows, we assume that the singlet interaction constant is $\alpha_{2}=-0.1\alpha_{1}$.

The corrections of the energies of the two polarization components of the condensate take into account both interaction constants:

\begin{eqnarray}
\Delta E^{+}  = \alpha _1 n_+ + \alpha _2 n_- \\
\Delta E^{-}  = \alpha _1  n_+ -\alpha _2 n_+
\end{eqnarray}

Where $E^{+,-}$  are the energy shifts of the $\sigma^+$ and $\sigma^-$ polarized component, $n_{+,-}$ are the exciton densities (including both the condensate and the reservoir) and $\alpha_{1,2}$, the interaction constants. If we decompose the total exciton density into reservoir density $n_r$ (the reservoir is not polarized) and condensate density $n_0$ (the condensate is polarized $\sigma^+$, as the pump), we can write the energy shifts as follows:

\begin{eqnarray}
\Delta E^{+} = \alpha _2 \frac{{n_r }}{2} + \alpha _1 \left(\frac{{n_r }}{2} + n_0 \right) \\
\Delta E^{-}  = \alpha _2 \left(\frac{{n_r }}{2} + n_0 \right) + \alpha _1 \frac{{n_r }}{2}
\label {EnergyShift}
\end{eqnarray}

Using this formula, we plot on the Figure 4 the energy shifts of the two components of the condensate as a function of the detuning, using the values of the reservoir density obtained by solving the full Boltzmann equations. The results presented on the figure correspond to a Rabi splitting of 4 meV and a temperature 5K.

 \begin{figure}[h]
  \includegraphics[width=0.5\textwidth,clip]{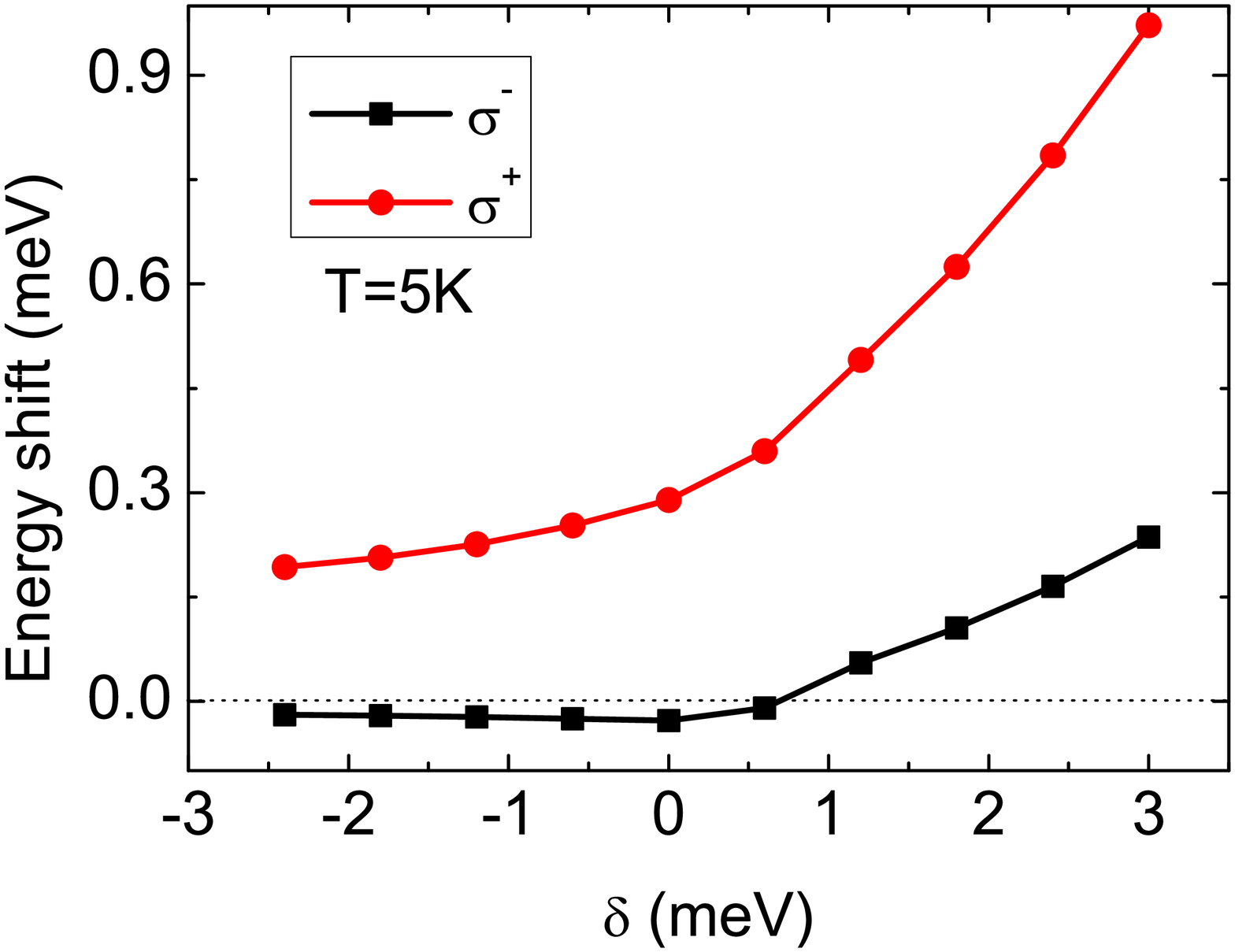}\\
  \caption{(Color online) Energy shift of $\sigma^+$ (red) and $\sigma^-$ (black) components of condensate . Rabi splitting 4 meV, temperature 5K.}
  \label{Fig4}
\end{figure}

We see that in the conditions of this simulation, namely, at relatively small Rabi splittings and quasi-resonant $cw$ pumping (as opposed to pulsed excitation), the contribution of the reservoir can totally modify the observed energy shifts. Indeed, at negative detuning the reservoir is not strongly populated, and a redshift of the $\sigma^-$ component is observed, as one would naturally expect from the negative sign of the $\alpha_2$ constant. However, at positive detuning the reservoir starts to play a role. Since the reservoir is depolarized, it contains a large number of $\sigma^-$ excitons which induce a blue shift for the $\sigma^-$ component of the condensate via the stronger repulsive interactions, which dominate the weak redshift induced by the $\sigma^+$ excitons.

Therefore, a measurement of the energy shift of the $\sigma^-$ component at positive detunings, without including the filling of long-lived reservoir states, can lead to inexact conclusions concerning the wrong sign and the magnitude of the polariton-polariton interaction constants.

\section{Application of the model}

One of the important effects based on the energy shifts of the polariton mode is the bistability (or multistability in the spinor case) of the polariton mode under quasi-resonant pumping. Indeed, if the system is pumped with a $cw$ laser whose frequency is above that of a bare polariton mode, blue shift of the macrooccupied mode will bring it closer to the laser, increasing the efficiency of the pumping. Therefore, at some pumping intensity the mode becomes unstable and the population jumps up abruptly. This effect has been predicted to occur also in the spinor case\cite{Gippius}, but if one takes into account the negative sign of $\alpha_2$, it is logical to expect that the jump of one polarization component under elliptical pumping will prevent the jump of the other component by moving it off-resonance.

In the first experimental work on polariton multistability\cite{Paraiso}, the authors have, in particular, observed an abrupt simultaneous increase of the amplitudes of both components, when the polarization of the excitation light is almost linear, but slightly elliptic (Fig 1(d,e) in Ref.\cite{Paraiso}). This contradicts the expected behavior mentioned above, and was explained by the authors by introducing a positive sign of $\alpha_2$, that is, repulsive interactions for excitons with opposite spins.

In this section we show how the formation of an excitonic reservoir in this particular type of experiment influences the observed behavior of the polariton modes in the multistable regime. In order to describe correctly the behavior of the macrooccupied modes under quasi-resonant pumping, we write the coupled Gross-Pitaevskii equations with pumping and decay for both polarization components $\Psi _ \pm$, taking into account the interactions with the reservoir.

\begin{widetext}
\begin{equation}
i\hbar \frac{{d\Psi _ \pm  }}{{dt}} = \left[ - i\frac{\gamma }{2} + \frac{\alpha _1}{S} |\Psi _ \pm  |^2  + \frac{\alpha _2}{S} |\Psi _ \mp  |^2  + \frac{{(\alpha _1  + \alpha _2 )}}{2}n_r \right]\Psi _ \pm   + P_ \pm  e^{i{\raise0.7ex\hbox{$\Delta $} \!\mathord{\left/
 {\vphantom {\Delta  \hbar }}\right.\kern-\nulldelimiterspace}
\!\lower0.7ex\hbox{$\hbar $}}t}
\label{GP}
\end{equation}
\end{widetext}

Here, $\gamma$ defines the broadening of the polariton line, $S$  the surface area of the polariton spot, $P_\pm$, the excitation density and $n_r$, the density of excitons in the reservoir. The detuning between the laser and the pumped polariton state $\Delta = 0.5 meV$, as in the experiment. The density of the reservoir can be obtained analytically, following the approach introduced in section I:

\begin{equation}
n_r  = \frac{1}{{2\pi }}\int\limits_0^\infty  {\frac{{N_{ph}^{k} \left(|\Psi _ +  |^2  + |\Psi _ -  |^2 \right)}}{{N_{ph}^{k}  + 1 + |\Psi _ +  |^2  + |\Psi _ -  |^2  + {\raise0.7ex\hbox{${\Gamma _k }$} \!\mathord{\left/
 {\vphantom {{\Gamma _k } {W_k }}}\right.\kern-\nulldelimiterspace}
\!\lower0.7ex\hbox{${W_k }$}}}}kdk}
\label{NRint}
\end{equation}

Taking an average value for $\frac {\Gamma _k } {W_k }$ and considering parabolic exciton dispersion, we can evaluate this integral analytically and obtain the solution for $n_r$ as a function of $\left(|\Psi _ +  |^2  + |\Psi _ -  |^2\right)$:

\begin{eqnarray}
 n_r  = \frac{1}{{4\pi }}\left(|\Psi _ +  |^2  + |\Psi _ -  |^2 \right)\frac{B}{C}\ln \left|1 + \frac{C}{A}\right|
\label{NRfin} \\
\nonumber A = \left[|\Psi _ +  |^2  + |\Psi _ -  |^2  + 1 + \frac{\Gamma }{W}\right]e^{\frac{{\delta ^{'} }}{{k_B T}}} \\
\nonumber B = \frac{{2m_{X} k_B T}}{{\hbar ^2 }} \\
\nonumber C =  - \left[|\Psi _ +  |^2  + |\Psi _ -  |^2  + \frac{\Gamma }{W}\right]
\label{NRfin2}
\end{eqnarray}

\begin{figure}[b]
  \includegraphics[width=0.5\textwidth,clip]{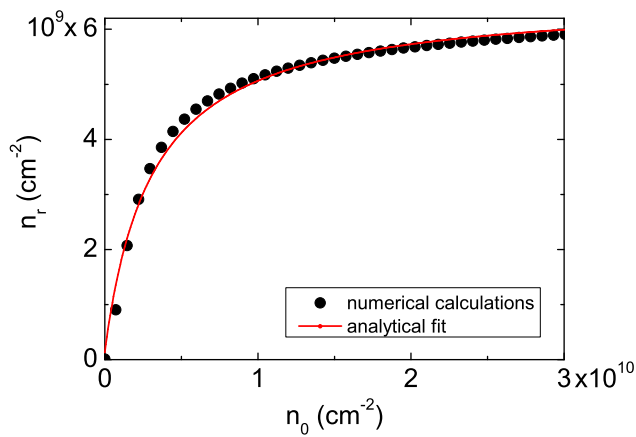}\\
  \caption{(Color online) Black circles - dependence of $n_r$ on $n_0$ calculated numerically. Red curve - fit of this data by the eqs. \ref{NRfin}. Fitting gives $\frac {\Gamma } {W } = 274$. This value corresponds to $n_{0}^{c}=\frac{N_{0}^{c}}{S}=3 \cdot 10^{9} cm^{-2}$}
  \label{Fig5}
\end{figure}

By introducing the above expression in the eq. \ref{GP}, we get a closed equation for the amplitude of the pumped polariton mode. An essential parameter of the model is the ratio $\frac {\Gamma } {W }$ (which has to be assumed constant in order to obtain analytical results). It can be estimated by calculating with the full system of semi-classical Boltzmann equations the value of the reservoir density $n_r$ as a function of the condensate density $n_{0}=\frac{|\Psi _ +  |^2  + |\Psi _ -  |^2}{S}$. We have then fitted the results obtained numerically by the analytical formula \ref {NRfin}. The results of this fitting are presented on Fig. 5. The value obtained for the fitting parameter $\frac {\Gamma } {W }$ was around 300. 

By changing the pump density slowly in time from 0 to some value and back again we can obtain the evolution of the populations of both circularly polarized polariton components and of the excitonic reservoir. The results are presented on the Fig.6. Indeed, a simultaneous jump of both circular components, and of the reservoir is visible when increasing the pumping power. The strong rise of the $\sigma^+$ polariton density associated with the passage of the bistable threshold leads to a strong rise of the population of the unpolarized excitonic reservoir which provokes a blue shift of the polariton energy of the $\sigma^-$ component. This blue shift overcomes by far the red shift induced by the attractive interaction between the opposite spin components of the condensate. It leads to the passage of the bistable threshold for the $\sigma^-$ component as well.

\begin{figure}[h]
  \includegraphics[width=0.5\textwidth,clip]{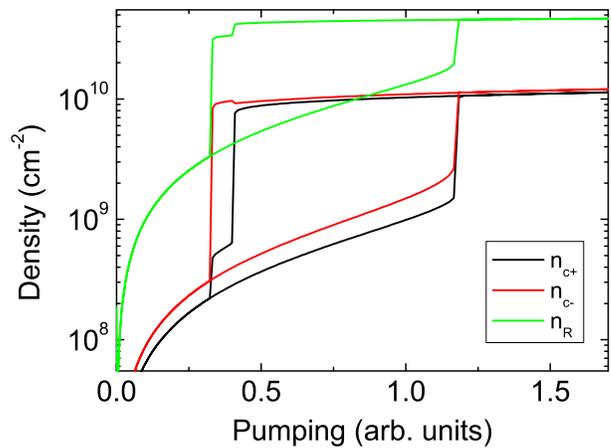}\\
  \caption{(Color online) Dependence of the amplitudes of $\sigma^+$ (red) and $\sigma^-$ (black) components on pump power. Polarization of pumping $\rho_{pump} = 0.1$}
  \label{Fig6}
\end{figure}

Another experimental effect is well reproduced when the system is on the upper stability branch and the pumping power is decreased. The $\sigma^-$-component has a smaller population and therefore jumps down first. At the same time, the population of the other component $\sigma^+$ strongly increases. This effect is easy to understand as well. When $\sigma^-$ jumps down, the energy of $\sigma^+$ is still higher that the laser frequency. The jump of $\sigma^-$ leads to a reduction of the total reservoir density $n_r$, bringing the frequency of the $\sigma^+$ component \emph{closer} to the laser frequency. This makes the pumping more efficient and increases the $\sigma^+$ population. When this amplitude in its turn abruptly decreases, the reservoir density jumps down once more, and this causes a second jump down of the $\sigma^-$-component. So, while the pumping power is decreased, the amplitude of $\sigma^-$-component undergoes two steps, as it was observed in experiment. 

However, in the framework of this model, we have not been able to fully describe the experimental behavior reported in \cite{Paraiso} for large circular polarization degree of the pump. We believe that the hypothesis of the fully unpolarized excitonic reservoir is no more valid in the case of quasi-circular pumping. Indeed, the jump up of one circular component only induces an energy splitting between $\sigma^-$ and 
$\sigma^+$ excitons in the reservoir, which may substantially slow down the spin relaxation. Thus, the reservoir may become spin polarized which would strongly affect the behavior of the pumped modes. The proper analysis of these phenomena requires a more complete description of the dynamics of the coupled condensate-reservoir system and will be addressed in future works.

The authors acknowledge the support of EU ITNs "Spin-Optronics" 237252 and "INDEX" 289968, IRSES "Polaphen" 246912 and ANR "Quandyde". We would like to thank Dr. Mikhail Glazov for useful discussions.

\section{Conclusion}

In this work we have studied the generation of an excitonic reservoir in the case of quasi-resonant pumping of a polariton mode. Exciton-polaritons from the condensate are scattered up with phonons. It was shown that in structures with small Rabi splitting and positive exciton-photon detuning, this process provides a strongly populated reservoir which influences the behavior of the resonantly pumped polariton modes. Because of the reservoir, it is possible to observe a blueshift of the polariton condensate component polarized opposite to excitation, while the exciton-exciton interaction constant for opposite spins remains negative (attractive interaction). An analytical model allowed us to describe the multistable behavior of the polariton system recently evidenced experimentally \cite{Sarkar,Paraiso}.

\end{document}